\documentstyle[twoside,epsf,epsfig]{article}

\catcode`\@=11
\long\def\@makefntext#1{
\protect\noindent \hbox to 3.2pt {\hskip-.9pt
$^{{\eightrm\@thefnmark}}$\hfil}#1\hfill}               %CAN BE USED

\def\@makefnmark{\hbox to 0pt{$^{\@thefnmark}$\hss}}    %ORIGINAL

\def\ps@myheadings{\let\@mkboth\@gobbletwo
\def\@oddhead{\hbox{}
\rightmark\hfil\eightrm\thepage}
\def\@oddfoot{}\def\@evenhead{\eightrm\thepage\hfil
\leftmark\hbox{}}\def\@evenfoot{}
\def\sectionmark##1{}\def\subsectionmark##1{}}

\oddsidemargin=\evensidemargin
\newcounter{sectionc}\newcounter{subsectionc}\newcounter{subsubsectionc}
\renewcommand{\section}[1] {\vspace{12pt}\addtocounter{sectionc}{1}
\setcounter{subsectionc}{0}\setcounter{subsubsectionc}{0}\noindent
        {\tenbf\thesectionc. #1}\par\vspace{5pt}}
\renewcommand{\subsection}[1] {\vspace{12pt}\addtocounter{subsectionc}{1}
\setcounter{subsubsectionc}{0}\noindent
{\bf\thesectionc.\thesubsectionc. {\kern1pt \bfit #1}}\par\vspace{5pt}}
\renewcommand{\subsubsection}[1] {\vspace{12pt}\addtocounter{subsubsectionc}{1}
        \noindent{\tenrm\thesectionc.\thesubsectionc.\thesubsubsectionc.
        {\kern1pt \tenit #1}}\par\vspace{5pt}}
\newcommand{\nonumsection}[1] {\vspace{12pt}\noindent{\tenbf #1}
        \par\vspace{5pt}}

\newcounter{appendixc}
\newcounter{subappendixc}[appendixc]
\newcounter{subsubappendixc}[subappendixc]
\renewcommand{\thesubappendixc}{\Alph{appendixc}.\arabic{subappendixc}}
\renewcommand{\thesubsubappendixc}
        {\Alph{appendixc}.\arabic{subappendixc}.\arabic{subsubappendixc}}

\renewcommand{\appendix}[1] {\vspace{12pt}
        \refstepcounter{appendixc}
        \setcounter{figure}{0}
        \setcounter{table}{0}
        \setcounter{lemma}{0}
        \setcounter{theorem}{0}
        \setcounter{corollary}{0}
        \setcounter{definition}{0}
        \setcounter{equation}{0}
        \renewcommand{\thefigure}{\Alph{appendixc}.\arabic{figure}}
        \renewcommand{\thetable}{\Alph{appendixc}.\arabic{table}}
        \renewcommand{\theappendixc}{\Alph{appendixc}}
        \renewcommand{\thelemma}{\Alph{appendixc}.\arabic{lemma}}
        \renewcommand{\thetheorem}{\Alph{appendixc}.\arabic{theorem}}
        \renewcommand{\thedefinition}{\Alph{appendixc}.\arabic{definition}}
        \renewcommand{\thecorollary}{\Alph{appendixc}.\arabic{corollary}}
        \renewcommand{\theequation}{\Alph{appendixc}.\arabic{equation}}  
        \noindent{\tenbf Appendix \theappendixc #1}\par\vspace{5pt}}
\newcommand{\subappendix}[1] {\vspace{12pt}
        \refstepcounter{subappendixc}
        \noindent{\bf Appendix \thesubappendixc. {\kern1pt \bfit #1}}
        \par\vspace{5pt}}
\newcommand{\subsubappendix}[1] {\vspace{12pt}
        \refstepcounter{subsubappendixc}
        \noindent{\rm Appendix \thesubsubappendixc. {\kern1pt \tenit #1}}
        \par\vspace{5pt}}

\newcommand{\textlineskip}{\baselineskip=13pt}
\newcommand{\smalllineskip}{\baselineskip=10pt}
        
\newcommand{\copyrightheading}[1]
        {\vspace*{-2.5cm}\smalllineskip{\flushleft
        {\footnotesize International Journal of Modern Physics D, #1}\\
        {\footnotesize \copyright\kern2pt World Scientific Publishing
         Company}\\
         }}
        
\def\abstracts#1#2#3{{
        \centering{\begin{minipage}{4.5in}\footnotesize\baselineskip=10pt

        \parindent=0pt #1\par
        \parindent=15pt #2\par
        \parindent=15pt #3
        \end{minipage}}\par}}
        
\newcommand{\bibit}{\nineit}

\renewenvironment{thebibliography}[1]
        {\frenchspacing
        \ninerm\baselineskip=11pt
         \begin{list}{\arabic{enumi}.}
        {\usecounter{enumi}\setlength{\parsep}{0pt}
         \setlength{\leftmargin 12.7pt}{\rightmargin 0pt}%FOR 1--9 ITEMS
         \setlength{\itemsep}{0pt} \settowidth
        {\labelwidth}{#1.}\sloppy}}{\end{list}}

\newcounter{itemlistc}
\newcounter{romanlistc}
\newcounter{alphlistc}
\newcounter{arabiclistc}

\newcommand{\fcaption}[1]{
        \refstepcounter{figure}
        \setbox\@tempboxa = \hbox{\footnotesize Fig.~\thefigure. #1}
        \ifdim \wd\@tempboxa > 5in
           {\begin{center}
        \parbox{5in}{\footnotesize\smalllineskip Fig.~\thefigure. #1}
            \end{center}}
        \else
             {\begin{center}
             {\footnotesize Fig.~\thefigure. #1}
              \end{center}}
        \fi}

\newcommand{\tcaption}[1]{
        \refstepcounter{table}
        \setbox\@tempboxa = \hbox{\footnotesize Table~\thetable. #1}
        \ifdim \wd\@tempboxa > 5in
           {\begin{center}
        \parbox{5in}{\footnotesize\smalllineskip Table~\thetable. #1}
            \end{center}}
        \else
             {\begin{center}
             {\footnotesize Table~\thetable. #1}
              \end{center}}
        \fi}

\def\@citex[#1]#2{\if@filesw\immediate\write\@auxout
        {\string\citation{#2}}\fi
\def\@citea{}\@cite{\@for\@citeb:=#2\do
        {\@citea\def\@citea{,}\@ifundefined
        {b@\@citeb}{{\bf ?}\@warning
        {Citation `\@citeb' on page \thepage \space undefined}}
        {\csname b@\@citeb\endcsname}}}{#1}}
        
\newif\if@cghi
\def\cite{\@cghitrue\@ifnextchar [{\@tempswatrue
        \@citex}{\@tempswafalse\@citex[]}}
\def\citelow{\@cghifalse\@ifnextchar [{\@tempswatrue
        \@citex}{\@tempswafalse\@citex[]}}
\def\@cite#1#2{{$\null^{#1}$\if@tempswa\typeout 
        {IJCGA warning: optional citation argument
        ignored: `#2'} \fi}}

\def\pmb#1{\setbox0=\hbox{#1}
        \kern-.025em\copy0\kern-\wd0   

       \kern.05em\copy0\kern-\wd0
        \kern-.025em\raise.0433em\box0}

\def\fnt#1#2{\footnotetext{\kern-.3em
        {$^{\mbox{\scriptsize #1}}$}{#2}}}

\def\fpage#1{\begingroup   
\voffset=.3in
\thispagestyle{empty}\begin{table}[b]\centerline{\footnotesize #1}
        \end{table}\endgroup}

\def\runninghead#1#2{\pagestyle{myheadings}
\markboth{{\protect\footnotesize\it{\quad #1}}\hfill}
{\hfill{\protect\footnotesize\it{#2\quad}}}}
\font\tenrm=cmr10
\font\tenit=cmti10
\font\tenbf=cmbx10
\font\bfit=cmbxti10 at 10pt
\font\ninerm=cmr9
\font\nineit=cmti9

\font\eightrm=cmr8

\def\qed{\hbox{${\vcenter{\vbox{                  %HOLLOW SQUARE
   \hrule height 0.4pt\hbox{\vrule width 0.4pt height 6pt
   \kern5pt\vrule width 0.4pt}\hrule height 0.4pt}}}$}}

\def\rfr#1{eq.(\ref{#1})}
\def\rfrs#1#2{eqs.(\ref{#1})-(\ref{#2})}
\def\Rfr#1{Eq.(\ref{#1})}
\def\Rfrs#1#2{Eqs.(\ref{#1})-(\ref{#2})}

\def\bb{\bibitem}
\def\eqi{\begin{equation}}
\def\eqf{\end{equation}}
\def\mtc#1{\mathcal{#1}}
\def\mc#1#2{\left[\matrix{#1 \cr #2\cr}\right]}

\def\mtc#1{\mathcal{#1}}
\def\ga{\gamma}
\def\d{\delta}

\def\th{\theta}

\def\p{\pi}
\def\r{\rho}

\def\t{\tau}

\def\f{\phi}

\def\ps{\psi}
\def\og{\omega}

\def\D{\Delta}

\def\O{\Omega}
\def\ci{\cos{i}}
\def\si{\sin{i}}
\def\vass#1{\left\vert\ #1 \right\vert}
\def\rp#1#2{{#1\over#2}}

\def\dert#1#2{\rp{d{#1}}{d{#2}}}

\def\lb#1{\label{#1}}

\def\grc{gravitomagnetic}

\def\nd{node}
\def\pg{perigee}
\def\mal{mean anomaly}
\def\cl{clock}
\def\eft{effect}
\def\mash{Mashhoon}

\def\gecc{G_{lpq}}
\def\fincl{F_{lmp}}
\def\gsue{\rp{\gecc}{e}\propto e^{\vass{q}-1}}
\def\rappe{\rp{1-e^2}{e}}

\def\unoe{1-e^2}
\def\unoee{\sqrt{1-e^2}}

\def\cpl{C^{+}_{lmf}}
\def\dw{\r_{w}}

\def\derf{\dert{\fincl}{i}}
\def\derg{\dert{\gecc}{e}}
\def\icombl{l-2p+q}
\def\icombb{l-2p}
\def\doto{\dot\O}
\def\dotog{\dot\og}

\def\ma{{\mtc{M}}}
\def\grp{\gecc\propto e^{\vass{q}}}

\begin{document}
\setlength{\textheight}{7.7truein}    %FOR 2ND PAGE ONWARDS

\runninghead{Satellite gravitational
orbital
perturbations and the gravitomagnetic clock effect} {Satellite gravitational
orbital
perturbations and the gravitomagnetic clock effect}

\normalsize\textlineskip
\thispagestyle{empty} 
\setcounter{page}{1}

%\copyrightheading{}             %{Vol.~0, No.~0 (1999) 000--000}

\vspace*{0.88truein}
   
\fpage{1}
\centerline{\bf SATELLITE GRAVITATIONAL
ORBITAL
PERTURBATIONS} 
\vspace*{0.035truein}
\centerline{\bf AND THE  GRAVITOMAGNETIC CLOCK EFFECT}
\vspace*{0.37truein}
\centerline{\footnotesize LORENZO IORIO}
\vspace*{0.015truein}
\centerline{\footnotesize\it Dipartimento Interateneo di Fisica dell' Universit{\`{a}} di
Bari, Via Amendola 173}
\baselineskip=10pt  
\centerline{\footnotesize\it Bari, 70126, Italy}  
\vspace*{0.225truein}
%\publisher{(received date)}

\vspace*{0.21truein}
\abstracts{In order to detect the \grc\ \cl\ \eft\ by means  
of two counter-orbiting satellites
placed on identical equatorial and circular orbits around the Earth 
with radius 7000 km their radial and azimuthal positions must be known
with an accuracy of $\d r
=10^{-1}$ mm and $\d \f=10^{-2}$ mas per revolution. In this work we investigate if the radial and azimuthal perturbations
induced by the dynamical and static parts of the Earth' s gravitational field  meet these requirements.
While the radial direction is affected only by harmonic perturbations with periods up to some tens of days, the azimuthal
location is perturbed by a secular drift and  very long period
effects.
% Concerning the time-varying
%part, it results that
%the
%radial perturbations, induced only
%by the odd degree tesseral ocean tides, have nominal
%amplitudes of the order of $10^1$ cm. Their periods are sufficiently short so that they can average out.
% but periods of some tens of days, so that they can average out over
%time spans smaller than the satellite' s lifetime.
%The situation is quite different for the azimuthal perturbations.
%They are caused by the solid and ocean even degree zonal tides. Among them the semisecular 18.6-year and 9.3-year tides
%stand out due to their relevant nominal amplitudes, amounting to $10^2 - 10^4$ mas, and very long periods. 
%Since the mission design requires relatively low altitudes for the
%satellites to be employed, their lifetimes, heavily affected by the
%atmospheric drag, could not allow to such long period tidal perturbations to average out because they could not be in time to
%describe a full cycle.
%Concerning the static part of the geopotential, the azimuthal perturbations are induced by the even zonal
%terms and the radial perturbations are generated by the odd tesseral terms. While the former ones are linearly growing in time,
%the latter ones have relatively short periods and average out.
It results that the present level of accuracy in the knowledge both of the Earth solid and ocean
tides, and of the static part
of the geopotential does not allow an easy detection of the
\grc\ \cl\ \eft\, at least by using short arcs only.}{}{}

\vspace*{1pt}\textlineskip      %) USE THIS MEASUREMENT WHEN THERE IS
\section{Introduction}    %) A SECTION HEADING
\vspace*{-0.5pt}
\noindent
Latest years have seen great efforts, both from a theoretical and an experimental point of
view, devoted to
the measurement of
the general relativistic Lense-Thirring effect$^{1,2}$
in the weak gravitational field
of the Earth by means of artificial satellites.

%Among the various gravitational \ct{kau} and non-gravitational [{\it Milani et al.,} 1987] perturbations which affect the orbital
%motion
%of near-Earth satellites those induced by the Earth solid and ocean tides play a relevant role. They can be modeled in terms
%of sinusoids with a wide interval of periodicities going from few days, depending on satellite, to the semisecular 9.3-year and
%18.6-year tides [{\it Balmino}, 1974; {\it Dow,} 1988; {Christodoulidis et al.}, 1988; {\it Casotto,} 1989; {\it Iorio,} 2000a; 2000b].
%The amplitudes depend from the satellite's orbital parameters: e.g. for LAGEOS they can reach the level of $10^3$ milliarcseconds
%(mas in the following). Their subtle action must be accurately accounted for in many of the presently designed experiments whose
%goal is to test, by means of artificial satellites, in the gravitational field of the Earth some features predicted by the \gr.
%More precisely, for a given tidal constituent the relation among the amplitude $A$, the period $P_{pert}$ and the
%observational time span $T_{obs}$ is very important since it could happen that a perturbation with a great $A$, which could 
%mask the general relativistic \eft\ investigated, 
%averages out
%on the chosen $T_{obs}$ because it is an integer multiple of its $P_{pert}$. It must be considered, in order that it
%realizes, it is necessary
%that $P_{obs}$ is as shorter as possible than the satellite' s lifetime; remember that only \lg-type satellites, almost not
%affected by the atmospheric drag, have lifetimes of the order of $5\cdot 10^5$ years.  

Among the satellite-based experiments recently proposed, one of the most
interesting is devoted to the detection of the  the \grc\ \cl\ \eft$^{3,4,5}$. It consists  in the
fact that two clocks moving along pro- and
retrograde circular equatorial
orbits, respectively, about the Earth exhibit a difference in
their proper times which, if calculated after some fixed angular
interval, say $2\p$, amounts
to:\eqi (\t_{+}-\t_{-})_{\f=2\p}\simeq 4\p\ \rp{J_{\oplus}}{M_{\oplus}c^2}\simeq 10^{-7}\ s,\eqf
where $J_{\oplus}$ and $M_{\oplus}$ are the
rotational angular momentum and the mass, respectively, of the Earth; $c$ is the speed of
light $in\ vacuo$.
In$^{4,6}$ it has been shown that 
for an orbit radius of 7000 km the radial
and azimuthal locations of the satellites must be known at a level of
accuracy of $\d r\leq
10^{-1}$ mm and $\d \f\leq 10^{-2}$ mas per revolution.
However, the studies conducted up to now on the feasibility
and the error budget of such an experiment are still preliminary$^{4,6}$.

In this paper we shall investigate in a quantitative manner  the
systematic errors induced on
the radial and azimuthal locations  by the Earth solid and ocean tidal perturbations and  by the static part of the 
geopotential. The paper is organized as follows.
In Section 2 and 3 the radial and azimuthal perturbations, respectively, induced by the most relevant tidal constituents are investigated.
In Section 4 the radial and azimuthal perturbations generated by the static part of the geopotential are worked out. Section 5 is devoted
to the conclusions. 
%---------------------------------------------

\section{The radial error induced by the Earth solid and ocean tides}
\noindent
According to$^{7,8}$ ,
the position perturbations in the radial direction can be expressed in general as:
\eqi \D r=\sqrt{{\D a}^2+\rp{1}{2}\ [\ (e\D a+a \D e)^2+(ae\D
\ma)^2]},\lb{drgen}\eqf where $a,\ e$ and $\ma$ denote the satellite' s semimajor axis,  eccentricity and  mean anomaly. 
In \rfr{drgen} the perturbation amplitudes are the rss values of the
perturbations and small eccentricity approximations have been extensively applied.

Since the difference in the proper orbital periods to be investigated is
integrated over 2$\p$ with respect to the azimuthal angle $\f$, as viewed by an
inertial observer fixed with the distant quasars, we shall consider only the
long period perturbations averaged over an orbital revolution.
This is accomplished by assuming those values for the indices $l,
p,\ q$$^{9,10}$ which
satisfy the relation $\icombl=0$. 
Since  the tidal perturbations on
the semimajor axis $a$ are proportional just to $\icombl$, all the terms in 
$(\D  
a)^2$ and $\D a$ of \rfr{drgen} vanish and it reduces to: 
\eqi \D r=\sqrt{\rp{1}{2}\ [\ (a \D e)^2+(ae\D
\ma)^2]}.\lb{drgen2}\eqf
For $e=0$ \rfr{drgen2} becomes:
\eqi  \D r_{tides}=\rp{1}{\sqrt{2}}\ a\ \D e_{tides}.\lb{dr}\eqf
For a constituent characterized by a given set of indices $\{l,\ m,\ p,\
q\}$, the first order tidal
perturbation amplitude for the eccentricity turns out to be\footnote{This
holds for $\icombl=0$. Note that for the eccentricity there are no second order, 
indirect perturbations due to the oblateness of the Earth$^{9,10,11}$, contrary to the
\nd, the \pg\ and the \mal.}: \eqi \D e_{lmpq}\propto
-\rp{(\icombb)\fincl\gecc}{e},\lb{eccper}\eqf
where $\fincl$ and $\gecc$ are the inclination and the eccentricity functions,
respectively, as can be found in$^{12}$.

\Rfr{eccper} allows to obtain a preliminary insight into those perturbations
which, for a given set of indices $l,\ m,\ p,\ q$, vanish.
Note that, for $e=0$, \rfr{eccper} could become singular. Concerning this problem
it must be considered that, since $\grp$, the behaviour of $\gsue$ is crucial.
If $\vass{q}-1\equiv k>0$, i. e. $q >1$ or $q< -1$, then for a circular orbit
$\rp{\gecc}{e}=0$ and the
perturbation vanishes. If $\vass{q}-1=0$, i.e. $q=\pm 1$, then 
$\rp{\gecc}{e}=const$.  Problems may arise only if $q=0$, but we shall see that, in general,
in
the cases in which $q$ takes such value,  $\icombb=0$ also holds so that 
the
perturbations identically vanish with no regards to the eccentricity or the
inclination of the satellite.

Let us start with the tidal perturbations of even degree. For $l=2n,\
p=0,...,l$ and $l-2p+q=0$,the allowable values for q satisfy the above
stated conditions so that we
can conclude that there are no radial tidal perturbations of even
degree. Since for the solid Earth tides we consider only the $l=2$ constituents,
this result rules out their possible influence on the radial error budget in the 
\grc\ \cl\ experiment.

Now we shall consider the odd degree case. For $l=3$ there are no problems
because 
$q$ never vanishes.
Moreover, for $p=1,\ q=-1$ and $p=2,\ q=1$ $\rp{\gecc}{e}=1$ and
$\icombb=\pm 1$ while for the other
sets of indices the perturbations vanish because $\rp{\gecc}{e}=0$. 
%For $l=5$ $\rp{\gecc}{e}$
%and $\icombb$ are not zero
%only for $p=2,\ q=-1$ and $p=3, q=+1$.
When we consider the inclination functions 
corresponding to the indices for which $\rp{\gecc}{e}$ and
$\icombb$ differ from zero, i. e. $F_{3m1}$ and $F_{3m2}$,  and evaluate them for
$i=0$
we
find that only
$F_{311}(i=0)=-\rp{3}{2}$.

So we can conclude that the radial direction is perturbed only by the 
$l=3,\ m=1,\ p=1,\ q=-1$ ocean tides.

The full expression for the eccentricity perturbation amplitude due to ocean
tides$^{9,10}$, in our case,
is given by\footnote{Here we shall consider only the prograde waves$^{13}$.}:
\eqi \D e_{lmpqf}=\rp{4\p G\dw}{na\dot\ga^{+}_{lmpqf}}\
(\rp{R_{\oplus}}{a})^{l+2}\
(\rp{1+k^{'}_{l}}{2l+1})\ \cpl\ [-\rp{(\icombb)\fincl\gecc}{e}],\lb{ecc}\eqf
where:\\
$\bullet\  G=6.67259\cdot 10^{-8}\ cm^{3}g^{-1}s^{-2}$ is the Newtonian
gravitational
constant$^{14}$\\
$\bullet\  n=\sqrt{\rp{GM_{\oplus}}{a^3}}$ is the satellite mean
motion; $GM_{\oplus}=3.986\cdot 10^{20}\ cm^{3}s^{-2}$ $^{14}$\\
$\bullet\  \dw=1.025\ g\ cm^{-3}$ is the water density$^{14}$\\
$\bullet\  R_{\oplus}=6378\cdot 10^{5}\ cm$ is the Earth' s equatorial radius\\
$\bullet\  k^{'}_{l}$ is the load Love number. $k^{'}_2=-0.3075, k^{'}_3=-0.1950$$^{15,16}$\\
$\bullet\  \cpl,\ cm$ is the ocean tidal height as can be found in EGM96 model$^{17}$\\
$\bullet\  \dot\ga^{+}_{lmpqf}=(\icombb)\dotog+(\icombl)\ma+m\doto+(j_2-m)\ \dot s+j_3\ \dot h+
j_4\ \dot
p+j_5\ \dot N^{'}+j_6\ \dot p_s$ in which $\og$ and $\O$ are the
satellite' s perigee and node and  the integers $j_i,\ i=2,..6$ refer
to the Doodson
number$^{18,19}$ by which each tidal constituent is classified. For
the astronomical
longitudes $s,...p_s$ see, e.g.$^{19}$. Recall that $\dot\ga^{+}_{lmpqf}$ has to be evaluated on the chosen reference orbit.\\
For $l=3,\ m=1,\ p=1,\ q=-1$, and by putting
$\dot\ga^{+}_{lmpqf}=\rp{2\p}{P_{pert}}$, \rfr{dr} becomes:
\eqi \D r_{311-1f}=(8.80\cdot 10^{25}\ cm^{7/2}s^{-1})\times a^{-7/2}\times
P_{pert}\times\cpl.\lb{eqper}\eqf
Among the tesseral tides, the $K_1$ (165.555) is by far the most important 
in perturbing the near Earth satellites'  orbits$^{20,21}$.
So it seems reasonable to calculate \rfr{eqper} for it in order to obtain an upper
bound in the order of magnitude of the tidally induced perturbations on $\D r$.
For such a constituent:\eqi C^{+}_{31}(K_1)=0.95\ cm,\lb{egm}\eqf
and
$\dot\ga^{+}_{311-1}(K_1)=
\dotog+\doto$. If we assume as reference orbit a secularly precessing ellipse$^{12}$, we obtain:
%for it the secular rates of the \nd\ and the \pg\ due to
%Earth' s oblateness \ct{kau}, which are not singular for $e=i=0$, we obtain:
\eqi
P_{pert}=-\rp{4\p}{3}\ 
\rp{1}{C_{2,0}R^2_{\oplus}\sqrt{GM_{\oplus}}}\ a^{7/2}=(4.7639\cdot
10^{-25}\ cm^{-7/2}s)\times a^{7/2}.\lb{k1}\eqf
\Rfrs{eqper}{k1} tell us the important feature that for $K_1,\ l=3,\ m=1,\ 
p=1,\ q=-1$ the perturbation amplitude is independent from the satellite'
s semimajor axis. For $a=7000$ km and $C_{2,0}=-J_{2}=-0.00108261$ we obtain
$P_{pert}=50$ days. By using \rfrs{egm}{k1} in \rfr{eqper} we obtain:
\eqi \D r_{311-1}(K_1)=39.849\ cm.\eqf
According to EGM96 model$^{17}$, the percentage error on
$C_{31}^{+}(K_1)$ amounts to $5.2\%$; this yields $\d r_{311-1}(K_1)\simeq 2.07$ cm. Despite
the amplitude of this long period mismodeled
perturbation is 2 orders of
magnitude
greater than the maximum allowable error $\d r_{max}=1\cdot 10^{-1}$ mm,
it must be noted that its
period $P_{pert}$ amounts to only 50 days. This implies that if an observational time span $T_{obs}$
which is an integer multiple of $P_{pert}$, i. e. some months, is adopted the tidal perturbative action
of $K_1$ can be averaged out.
%Moreover, the
%magnitude of the perturbation induced seems to rule out the possibility of
%using an $a=7000$ km Earth satellite in order to detect the \grc \cl\ \eft.
%--------------------------------------------------

\section{The azimuthal error induced by the Earth solid and ocean tides}
\noindent
Concerning the angular variable which defines the position of the satellite on the orbit,
for an equatorial, circular orbit it seems reasonable to adopt for its rate of change:
\eqi\dot\f=\dot\og+\dot\O\ci+\dot\ma.\lb{meanlong}\eqf In it $\dot\og+\dot\O\ci$ represents an angular
velocity around
the direction of the orbital angular
momentum$^{22}$; it is valid
for any inclination angle $i$. In order
to account for the fact that the orbit is
circular we add to it $\dot\ma$. See also$^{23}$.
About the perturbations on the latter Keplerian orbital element, it turns out
that$^{22}$
in $\D \ma$ one has to consider also the indirect perturbations on the 
mean motion $n$ due to the cross coupling with the semimajor axis $a$. Since they are 
proportional to $\icombl$, they vanish when only long period perturbations are considered,
as is the case here. 
The perturbation amplitudes on the \nd, the \pg\ and the \mal\
are proportional to\footnote{The second order, indirect perturbations will not be considered here since
it can be 
demonstrated that they vanish in this case.}:
\eqi \D\O_{lmpqf}\propto \rp{1}{\unoee\si}\ \gecc\dert{\fincl}{i},\lb{nodo}\eqf
\eqi \D\og_{lmpqf}\propto \rp{\unoee}{e}\ \derg\ \fincl-\rp{\ci}{\unoee\si}\
 \gecc\ \derf,\lb{perig}\eqf
\eqi \D\ma_{lmpqf}\propto -\rappe\ \derg\ \fincl+2(l+1)\ \fincl\ \gecc.\lb{mean}\eqf
By assuming $\unoe=\unoee\simeq 1$ for $e\rightarrow 0$, with the aid of \rfrs{meanlong}{mean}
we obtain:
\eqi\D\f_{lmpqf}\propto 2(l+1) \fincl\ \gecc.\lb{dfvera}\eqf
As already done in the previous section, \rfr{dfvera} can be used in order to forecast
which perturbations will vanish.

For $l=2$ only the combination $l=2,\ p=1,\ q=0$ yields a nonzero eccentricity
function: $G_{210}=(\unoe)^{-3/2}=1$. Among the corresponding inclination functions $F_{2m1}$,
for $i=0$ we have $F_{201}=-1/2$. The same conclusion holds also for $l=4$ with
$G_{420}(e=0)=1$ and $F_{402}(i=0)=3/8$.
So we can conclude that for $l=2,4$ only the zonal tides, both solid and ocean, cause
nonvanishing perturbations on the satellite' s azimuthal variable.

Concerning the odd degree perturbations, they all vanish since for $l=3,5$, $q$
is always nonzero, so that, since $\grp$, for circular orbits all the eccentricity
functions vanish. The conclusion is that the odd part of the ocean tidal spectrum
does not induce systematic errors on the satellite's azimuthal variable.

For a given constituent of degree $l$, order $m$ and frequency $f$ the full 
expressions for the solid
Earth and ocean tidal perturbation amplitude (progressive waves only) are, respectively:
\eqi \D \f^{solid}_{lmpqf}=\rp{g}{n a^2 \dot\ga_{lmpqf}}\
(\rp{R_{\oplus}}{a})^{l+1}\ A_{lm}\ k^{(0)}_{lmf}\ H^{m}_{l}
\ [2(l+1)\ \fincl\ \gecc],\lb{azis}\eqf 
\eqi \D \f^{ocean}_{lmpqf}=\rp{4\p G\dw}{na\dot\ga^{+}_{lmpqf}}\
(\rp{R_{\oplus}}{a})^{l+2}\
(\rp{1+k^{'}_{l}}{2l+1})\ \cpl\ [2(l+1)\ \fincl\ \gecc],\lb{azioc}\eqf
where:\\
$\bullet\ g=978.0327\ cm\ s^{-2}$  is the acceleration of gravity at the surface of
the Earth as if it was 
perfectly spherical$^{14}$\\
$\bullet\ \dot\ga_{lmpqf}=\dot\ga^{+}_{lmpqf}$\\ 
$\bullet\ A_{lm}=\sqrt{\rp{2l+1}{4\p}\rp{(l-m)!}{(l+m)!}}$\\
$\bullet\ k^{(0)}_{lmf}$ is the Love number for the free space potential$^{14,24}$\\
$\bullet\ H^{m}_{l}$ are the Doodson coefficients with a different normalization$^{14,25}$\\
We shall start by considering the three most relevant $l=2\ m=0$ zonal
tides:\\
$\bullet\ 18.6-year\ (055.565);\ P_{pert}=6798.38\ 
days;\ k^{(0)}_{20}=0.315;\ H^0_2=2.792\
cm$\\
$\bullet\ 9.3-year\ (055.575);\ P_{pert}=3399.19\ 
days;\ k^{(0)}_{20}=0.313;\ H^0_2=2.72\cdot 10^{-2}\
cm$\\
$\bullet\ S_{a}\ (056.554);\ P_{pert}=365.27\ 
days;\ k^{(0)}_{20}=0.307;\ H^0_2=-4.92\cdot 10^{-1}\
cm;\ k^{'}_{2}=-0.3075;\ \cpl=2.54\ cm$\\
For $l=2$ \rfrs{azis}{azioc} becomes:
\eqi \D\f^{solid}=(-3.77\cdot 10^{18}\ cm^{5/2}s^{-1}) \times
a^{-7/2}\times
k^{(0)}_{20}\times P_{pert}\times H^{0}_{2},\lb{cazs}\eqf
\eqi \D\f^{ocean}=(-4.707\cdot 10^{17}\ cm^{5/2}s^{-1}) \times
a^{-7/2}\times
P_{pert}\times\cpl.\lb{cazzarol}\eqf
Note that, since for the $l=2$ zonal tides $P_{pert}$ does not
depend on the satellite' s semimajor axis but only on the astronomical
arguments, $\D\f_{lmpqf}$ depends on the orbit' s radius through
$a^{-7/2}$, contrary to $\D\ r(K_1)$, as shown in the previous
Section. For $a=7000$ km we have:\\
$\bullet\ \D\f(18.6-year)=-4.431\cdot 10^{4}$ mas\\
$\bullet\ \D\f(9.3-year)=-214.4$ mas\\
$\bullet\ \D\f(S_a)=408$ mas (solid); 857.6 mas (oceanic)

The zonal tidal perturbations on the satellite' s azimuthal location are
particularly insidious not only because their nominal amplitudes are up to 6
orders of magnitude greater than the maximum allowable error
$\d\f_{max}=10^{-2}$ mas, but also
because they have periods very long, so that there is no hope they average out
on reasonable $T_{obs}$. Concerning the 18.6-year tide, by
assuming an uncertainty
of $1.5\%$ on $k^{(0)}_{20}$$^{20}$, the mismodeling
on its perturbation amounts to 
-664 mas which is, however, very far from $\d\f_{max}$.
% Moreover, in order to maximize the 
%allowable error [{\it Mashhoon et al.,} 1998]
%$\rp{\d r}{r}\simeq\rp{\d\f}{\f}\simeq\rp{nJ_{\oplus}}{Mc^2}$,  relatively low
%altitude satellites are requested;
%this implies they are strongly affected by the atmospheric drag, so that 
%their lifetime could not allow to the semisecular tidal perturbations
%to describe a cycle at least. 
%---------------------------------

\section{Static geopotential perturbations}
\noindent
As can be found in$^{12}$, the perturbing function of degree $l$ and order $m$
of the static part of the geopotential can be cast into the form:
\eqi V_{lm}=\rp{GM_{\oplus}{R_{\oplus}}^l}{a^{l+1}}\ \fincl\ \gecc\ S_{lmpq},\lb{kaul}\eqf
where:\\
$\bullet\ S_{lmpq}={\mc{C_{lm}}{-S_{lm}}}^{l-m\ even}_{l-m\ odd}
\cos{\ps_{lmpq}}+{\mc{S_{lm}}{C_{lm}}}^{l-m\ even}_{l-m\ odd}\sin{\ps_{lmpq}}$\\
$\bullet\ C_{lm},\ S_{lm}$ are the unnormalized Stokes' geopotential coefficients$^{17}$\\
$\bullet\ \ps_{lmpq}=(\icombb)\og+(\icombl)\ma+m(\O-\th)$ in which $\th$ is the sidereal angle

Concerning the preliminary analysis of the  even degree perturbations, the same conclusions of
the previous sections hold. All the perturbations on the radial directions  vanish, $\D r_{static}=0$,
while for the satellite' s azimuthal location only the zonal contributions are to be considered.
Let us work out explicitly the perturbation due to the main even zonal coefficient $C_{2,0}$. 
For the precessional secular rates induced by it of a satellite on circular orbit we have:
\eqi \dert{\og}{t}=\rp{3}{4}\ \rp{nC_{2,0}R_{\oplus}}{a^2}\ (1-5\ci^2),\lb{grg1}\eqf
\eqi \dert{\O}{t}=\rp{3}{2}\   \rp{nC_{2,0}R_{\oplus}}{a^2}\ \ci,\eqf
\eqi \dert{\ma}{t}= -\rp{3}{4}\ \rp{nC_{2,0}R_{\oplus}}{a^2}\\ (3\ci^2-1).\lb{grg3}\eqf
By inserting \rfrs{grg1}{grg3}, evaluated for i=0, into \rfr{meanlong} we
obtain:
\eqi \dert{\f}{t}=-3\sqrt{GM_{\oplus}}C_{2,0}R^2_{\oplus}a^{-7/2}=(2.637\cdot
10^{25}\ cm^{7/2}s^{-1})\times
a^{-7/2}.\eqf
For a circular orbit of radius 7000 km the  azimuthal secular rate is:\eqi \dert{\f}{t}=1.89\cdot 10^{10}\ mas/y.\eqf
It is important to evaluate the error induced on such rate by the poor knowledge of the Earth' s gravitational
field. According to the EGM96 model$^{17}$, a relative uncertainty
of $7.3\cdot 10^{-8}$ 
weighs on the $C_{2,0}$ coefficient. This yields
$ \d\dot\f\simeq 1380$ mas/y, which is equivalent to $\d\f\simeq 2.5\cdot 10^{-1}$ mas per revolution,
being $P_{orb}=\rp{2\p}{n}=5.82\cdot 10^{3}$ s for $a=7000$ km. Such error is 1 order of magnitude greater than 
$\d\f_{max}\simeq 10^{-2}$ mas. 

Concerning the perturbations of odd degree, let us consider in detail the most important geopotential harmonic
of degree $l=3$.
By reasoning as in the previous
sections, it results that
the satellite' s azimuthal location is not perturbed  by the $l=3$ part of the geopotential spectrum.
Concerning $\D r$, the perturbation corresponding to the combination $l=3,\ m=1,\ p=1,\ q=-1$ does not
vanish.
The full expression for the geopotential perturbation on the eccentricity is given by:
\eqi \D e_{lmpq}=\rp{GM_{\oplus}}{na^2\dot\ps_{lmpq}}\ (\rp{R_{\oplus}}{a})^{l}\ \rp{\gecc}{e}\ \fincl\ [-(l-2p)]\ S_{lmpq}.\eqf
In our case it yields:
\eqi \D r_{static}=\rp{3}{2\sqrt{2}}\ \sqrt{GM_{\oplus}}R_{\oplus}^3
a^{-7/2}\times\rp{P_{pert}}{2\p}\times[C_{3,1}\cos{\ps_{311-1}}+S_{3,1}\sin{\ps_{311-1}}],\eqf
with:\eqi\dot\ps_{311-1}\equiv\rp{2\p}{P_{pert}}=\dotog+\doto-\dot\th.\eqf
With $a=7000$ km we have for the period and the amplitudes of the harmonic terms:
\eqi P_{pert}=8.81\cdot 10^{4}\ s,\eqf
\eqi A_{\D r}=(-8.481\cdot 10^9\ cm)\times\mc{C_{3,1}}{S_{3,1}}.\lb{pert3}\eqf
According to$^{17}$ the mismodeling weighing on the Stokes' coefficients of interest
amounts to $\d C_{3,1}=1.5\cdot 10^{-10},\ \d S_{3,1}=1.3\cdot 10^{-10}$. This leads to a mismodeled
radial perturbation:
\eqi \d r_{static}\simeq (1.3\ cm)\cos{\ps_{311-1}}+(1.15\ cm)\sin{\ps_{311-1}}.\eqf
It results to be 2 orders of magnitude greater than the allowable $\d
r_{max}\simeq 10^{-1}$ mm.
However, it must be pointed out that such a mismodeled perturbation averages out 
on an observational time span $T_{obs}$ which is an integer multiple of 1 day,  its
period amounting
to 24.48 hr.
The same conclusions can be drawn for the other higher odd degree nonvanishing radial perturbations.   
%------------------------------------------

\section{Discussion and conclusions}
\noindent
In this paper we have explicitly calculated, by averaging over one orbital revolution,  the most relevant perturbations $\D r$ and $\D \f$ due
to the dynamical and
static part of the Earth's gravitational field on the radial and azimuthal locations of a satellite
placed in an equatorial,
circular orbit with radius of 7000 km. Furthermore, we have  compared the mismodeling induced on such perturbations by
the poor knowledge of the parameters of the Earth' s gravitational field to the maximum  errors per revolution
$\d r_{max}\simeq 10^{-1}$
mm
and $\d\f_{max}\simeq 10^{-2}$ mas allowable in order to detect
successfully the \grc\ \cl\ \eft.

Concerning the radial direction, it is affected by harmonic perturbations induced by the odd degree part, both static
and dynamical,  of the Earth's gravitational field. If, from one hand, the related mismodeling is  2 orders of magnitude
greater than $\d r_{max}$, from the other hand it must be
pointed out that such mismodeled perturbations average
out on not too long time spans since their periods range from 1 to 50 days.
 
The situation for the azimuthal  angle is different. It is acted upon by the even degree zonal
harmonics of
the Earth' s gravitational field. The zonal tides are very insidious since they induce perturbations with
great amplitudes acting on very long periods, so that it is necessary
to wait for several years in order to average
out their mismodeled effect which are up to 4 orders of magnitude greater
than $\d\f_{max}$. The $l=2,\ m=0$ part of the geopotential
induces
also a secular drift on the azimuthal satellite' s
 angle; the uncertainties in $C_{2,0}$  induces on it a mismodeled rate per revolution
which is 1 order of magnitude grater than $\d\f_{max}$.

\begin{figure}[htbp]
\vspace*{13pt}
\centerline{\psfig{file=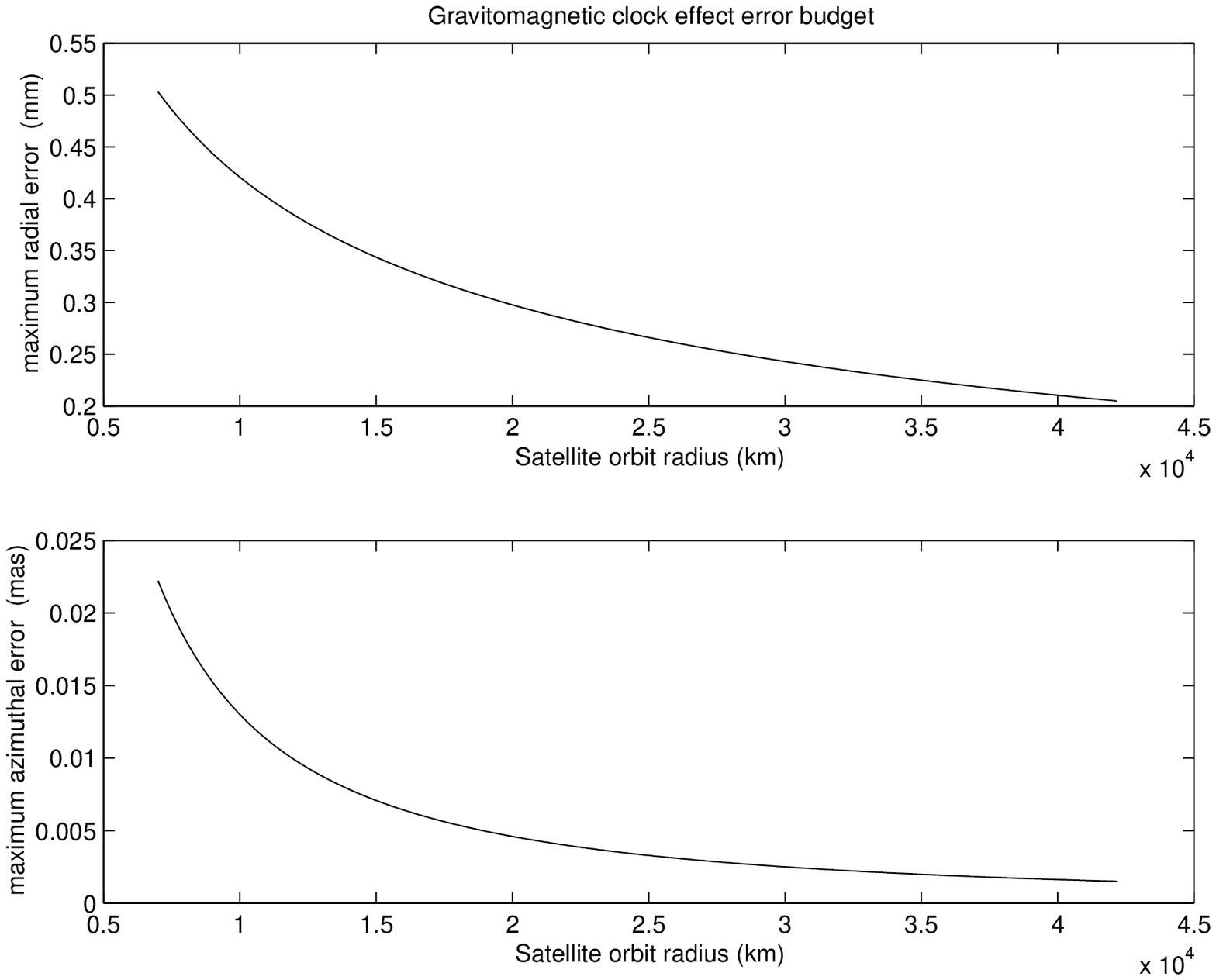,width=9.2cm}}
\vspace*{13pt}
\fcaption{Maximum allowable errors in the radial and azimuthal locations.
The values for the orbit radius span from 7000 km to 42160 km for a geostationary satellite.}
\label{errfig}
\end{figure}

\begin{figure}[htbp] %ORIGINAL SIZE: width=1.4TRUEIN; height=1.5TRUEIN
\vspace*{13pt}
\centerline{\psfig{file=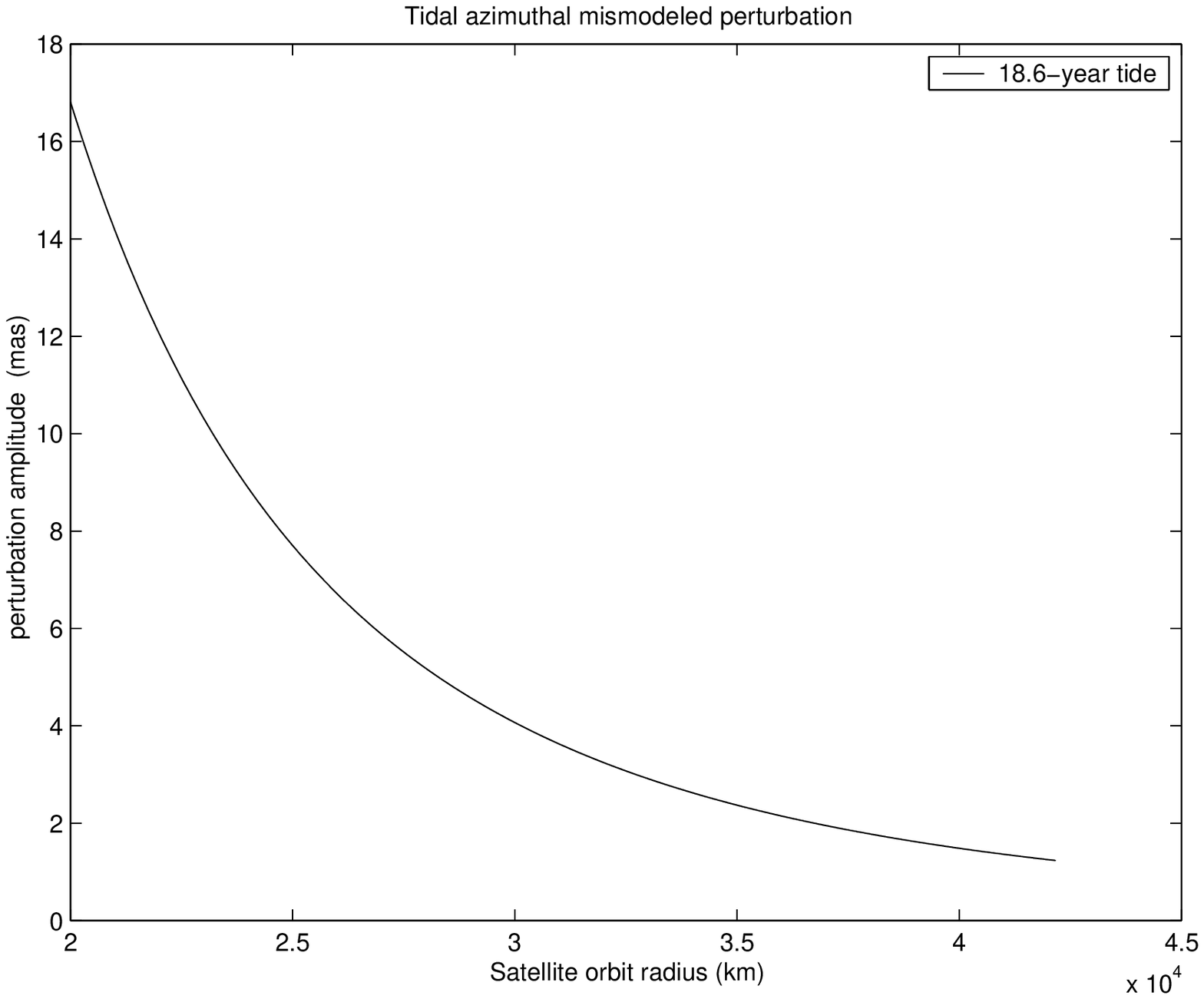,width=9.2cm}}
\vspace*{13pt}
\fcaption{Mismodeled azimuthal perturbation induced by the 18.6-year tide.}
\label{unofig}
\end{figure}

\begin{figure}[htbp]
\vspace*{13pt}
\centerline{\psfig{file=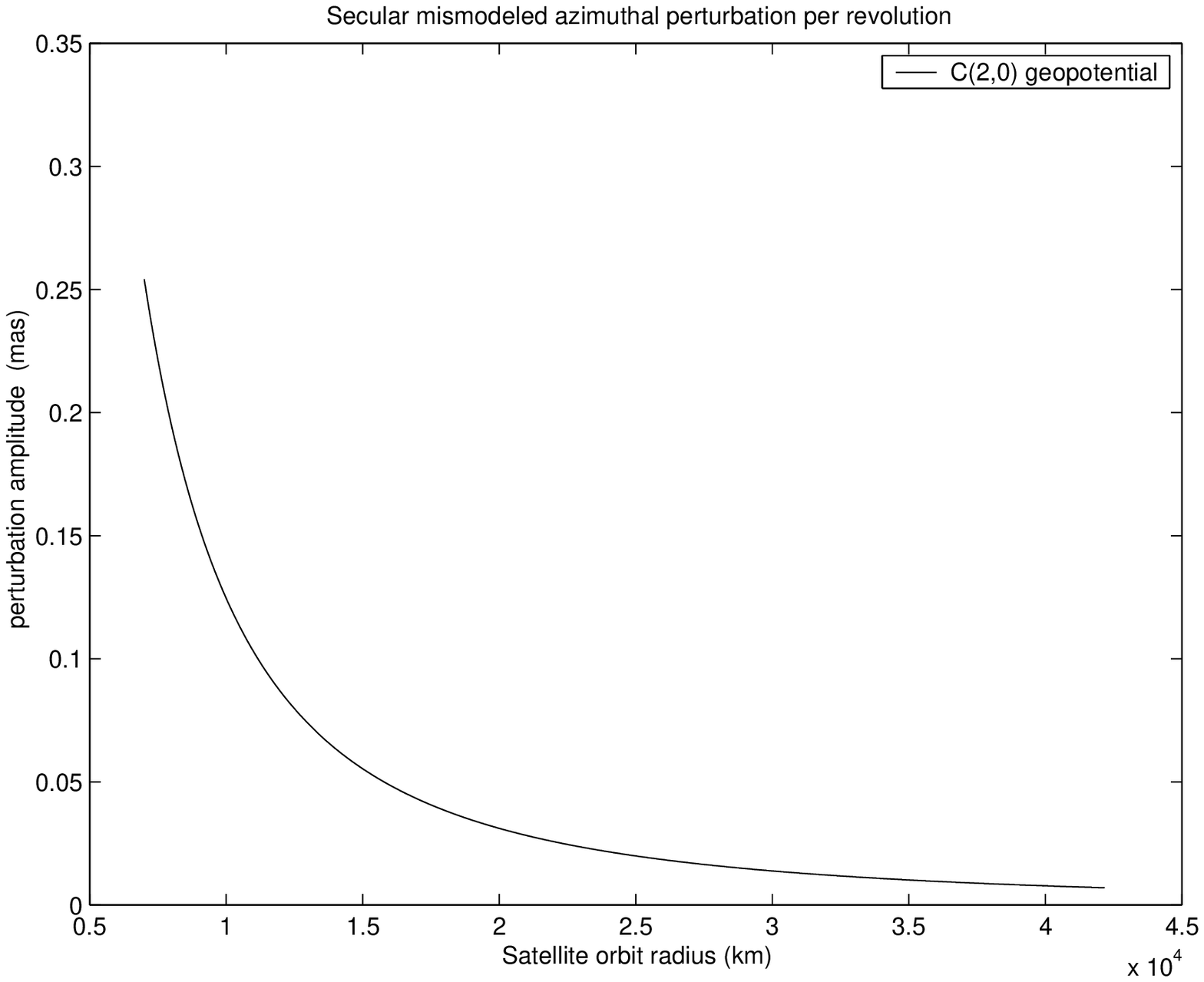,width=9.2cm}}
\vspace*{13pt}
\fcaption{Mismodeled azimuthal rate per revolution induced by the $C_{2,0}$ geopotential
coefficient.
The values for the orbit radius span from 7000 km to 42160 km for a geostationary satellite.}
\label{duefig}
\end{figure}

The conclusions outlined here hold for $r=7000$ km; let us see how the situation changes with 
different values for the orbital radius.
The possible scenarios turn out to be  very intricate. Indeed, from one hand we have secular or semisecular mismodeled
perturbations which could be reduced only 
by enlarging the orbit radius, and from the other hand there are the periodic mismodeled perturbations whose periods grow with the
orbit radius making so much more difficult to average them out  on reasonable
time spans.  Moreover, it should also be considered that 
the maximum allowable errors depend on the orbit radius
and they decrease with increasing radius putting, in this way, more
stringent
constraints on the mismodeled gravitational perturbations. This is shown
in Fig.(\ref{errfig}).
%--------------------------------
%-------------------------------------- 
In Fig.(\ref{unofig}) and Fig.(\ref{duefig}) we show how the mismodeled perturbative amplitudes due to the 18.6-year tide and the $C_{2,0}$
depend on the orbit radius.
%--------------------------------
%-------------------------------- 
%-------------------------------
From an inspection of Fig.(\ref{errfig}) and Fig.(\ref{unofig}) it can be noted that the major problems come from the azimuthal error
and the perturbation induced by the 18.6-year tide: as the orbit radius grows, the mismodeled tidal perturbation is always greater
than the 
maximum allowable error by 2 or 3 orders of magnitude. Since this important source of systematic error cannot be made harmless by
varying the orbital parameters of the
satellites, it should be necessary to average out its effect: but this means to choose a time span $T_{obs}=18.6$ years at least.

\begin{figure}[htbp]
\vspace*{13pt}
\centerline{\psfig{file=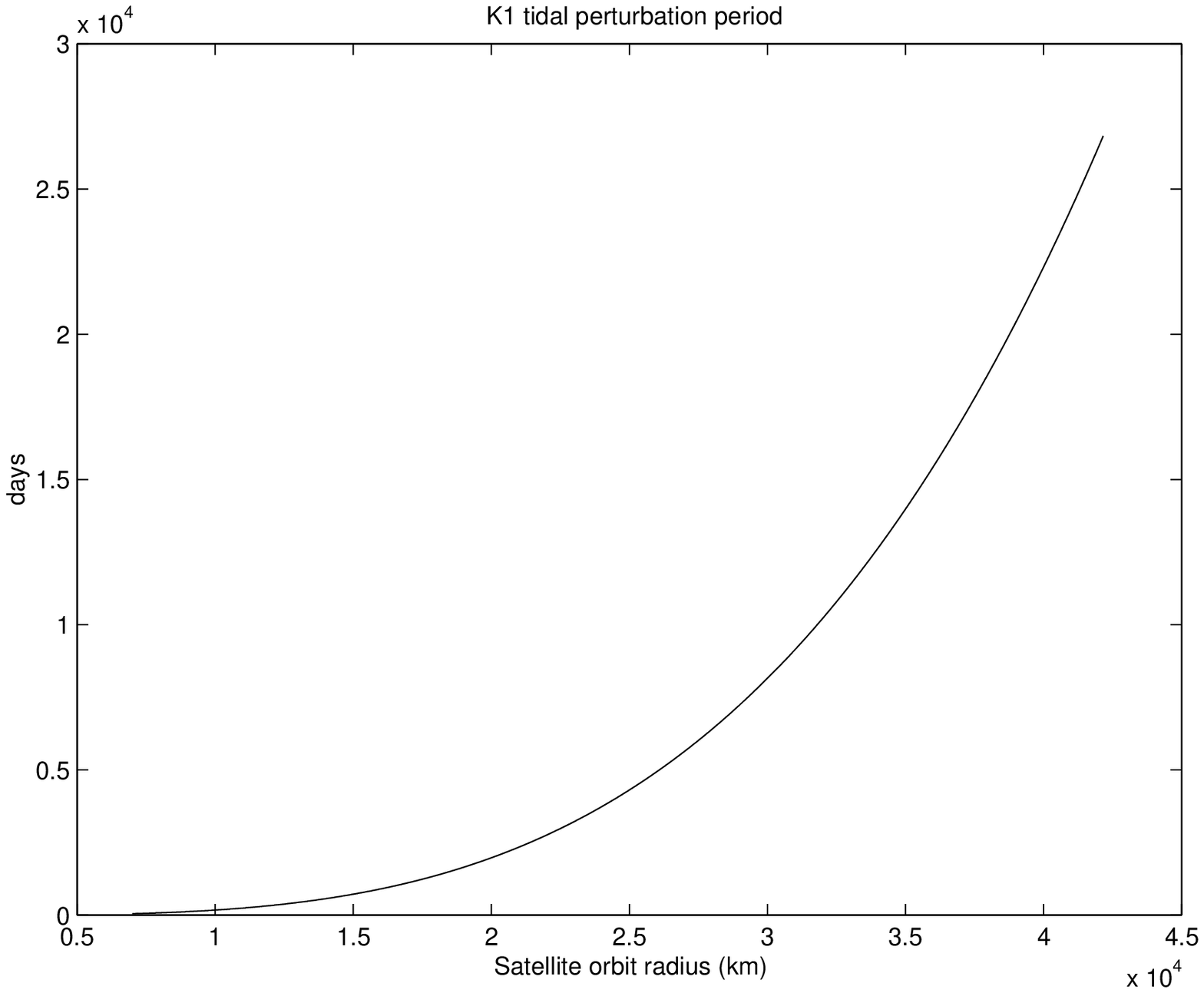,width=9.2cm}}
\vspace*{13pt}
\fcaption{Period of the $K_1\ l=3\ p=1\ q=-1$ radial tidal perurbation.
The values for the orbit radius span from 7000 km to 42160 km for a geostationary satellite.}
\label{perk1fig}
\end{figure}

By
inspecting Fig.(\ref{perk1fig}) it can be noted the growth of the period of the radial tidal perturbation 
is induced by the
$K_1,\ l=3\ p=1\ q=-1$. This  is an important feature since its mismodeled amplitude is at cm level and
is independent from the orbit
radius. Moreover, $\d r_{max}$ is of the order of $10^{-1}$ mm, so that we
could eliminate the effect of such a perturbation only by
averaging it over an integer multiple of its period.
%--------------------------------
%------------------------------- 

We can conclude that the present level of accuracy in the knowledge both of the Earth solid and ocean tides, and of the static part
of the geopotential does not allow
an easy detection of the \grc\ \cl\ \eft\, at least by using short arcs only.

\nonumsection{Acknowledgements}
\noindent
I wish to thank B. Mashhoon for the useful material supplied to me and for his useful suggestions.
I am also grateful to I. Ciufolini
and E. Pavlis. Special thanks  to L. Guerriero who supported me at Bari.

%---------------------------------
\nonumsection{References}

%*****************************
\end{document}